%% file: main.tex
\documentclass[%
 reprint,
 superscriptaddress,
 amsmath,amssymb,
 aps,
pra,
floatfix,
]{revtex4-2}

\usepackage{xcolor}
\usepackage{graphicx}
\usepackage{dcolumn}
\usepackage{bm}
\usepackage{diagbox}
\usepackage{braket} 
\usepackage{array}

\makeatletter
\newcommand{\thickhline}{%
    \noalign {\ifnum 0=`}\fi \hrule height 1pt
    \futurelet \reserved@a \@xhline
}
\newcolumntype{"}{@{\hskip\tabcolsep\vrule width 1pt\hskip\tabcolsep}}
\makeatother

\newcommand{\DavisECE}{Department of Electrical and Computer Engineering, University of California, Davis, CA 95616}
\newcommand{\DavisPhysics}{Department of Physics, University of California, Davis, CA 95616, USA}
\newcommand{\BerkeleyQNL}{Quantum Nanoelectronics Laboratory, Department of Physics, University of California at Berkeley, Berkeley, CA 94720, USA}
\newcommand{\BerkeleyComputation}{Computational Research Division, Lawrence Berkeley National Laboratory, Berkeley, California 94720, USA}

\newcommand{\LincolnLab}{Lincoln Laboratory, Massachusetts Institute of Technology, Lexington, MA 02421-6426}

\begin{document}

\preprint{APS/123-QED}

\title{Observation of Photon Blockade in a Tavis-Cummings System}

\author{Brian Marinelli}
\thanks{These authors contributed equally.}
\affiliation{\BerkeleyQNL}
\affiliation{\BerkeleyComputation}

\author{Alex H. Rubin}
\thanks{These authors contributed equally.}
\affiliation{\DavisECE}
\affiliation{\DavisPhysics}

\author{Victoria A. Norman}
\affiliation{\DavisECE}
\affiliation{\DavisPhysics}

\author{Santai Yang}
\affiliation{\BerkeleyQNL}

\author{Ravi Naik}
\affiliation{\BerkeleyQNL}
\affiliation{\BerkeleyComputation}

\author{Bethany M. Niedzielski}
\affiliation{\LincolnLab}

\author{David K. Kim}
\affiliation{\LincolnLab}

\author{Rabindra Das}
\affiliation{\LincolnLab}

\author{Mollie Schwartz}
\affiliation{\LincolnLab}

\author{David I. Santiago}
\affiliation{\BerkeleyQNL}
\affiliation{\BerkeleyComputation}

\author{Christopher Spitzer}
\affiliation{\BerkeleyQNL}
\affiliation{\BerkeleyComputation}

\author{Irfan Siddiqi}
\affiliation{\BerkeleyQNL}
\affiliation{\BerkeleyComputation}

\author{Marina Radulaski}
\thanks{Corresponding author: mradulaski@ucdavis.edu}
\affiliation{\DavisECE}

\date{\today}

\begin{abstract}
We observe blockade of microwave photons in a Tavis-Cummings system comprising a superconducting cavity and up to $N=3$ transmon qubits.
The effect is characterized with photon number-resolving spectroscopy using an additional dispersively coupled transmon ``witness" qubit to directly probe the cavity's photon number distribution.
We first observe polariton formation with splitting proportional to $\sqrt{N}$, confirming the Tavis-Cummings coupling, and subsequently obtain sub-Poissonian cavity photon statistics when the cavity is driven at polariton frequencies.
\end{abstract}

\maketitle

\section{Introduction}

\input{intro}

\begin{figure}[htp]
    \centering
    \includegraphics[width=\columnwidth]{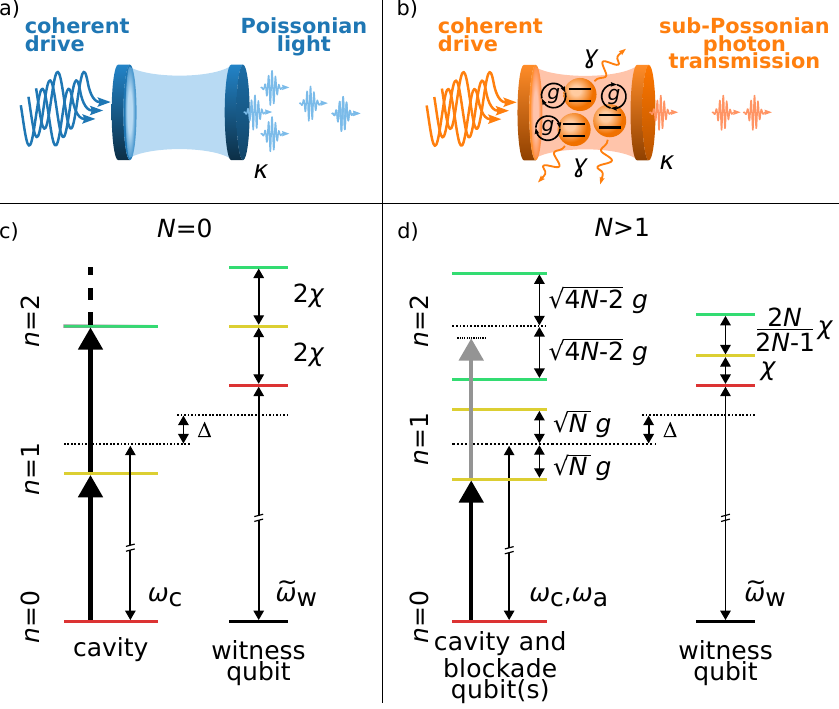}
    \caption{
    (a) Illustration of an optical cavity driven by a coherent field. The photons transmitted through the cavity have Poissonian statistics, maintaining the coherent nature of the drive field.
    (b) Illustration of an open Tavis-Cummings system with 3 two-level emitters coupled to a cavity mode. The photons transmitted through the cavity have sub-Poissonian statistics due to photon blockade.
    (c) First few energy levels of an empty electromagnetic cavity (left) and a ``witness" qubit dispersively coupled to the cavity (right). Dashed lines correspond to bare states. The witness qubit frequency shifts upward by $2\chi$ for each photon in the cavity. Spectroscopy of the witness qubit therefore reveals the cavity photon number distribution. Thick arrows represent a coherent pump resonant with the first excited state of the cavity and therefore also with all higher transitions.
    (d) Left: Highest and lowest levels of the first few excited-state manifolds of a system of $N$ identical qubits resonantly coupled to a cavity. Right: the corresponding energy levels of a witness qubit dispersively coupled to the same cavity. Dashed lines correspond to bare states.
    Thick arrows represent a coherent pump resonant with the lowest excited state, which is detuned from states in the $n=2$ manifold, giving rise to photon blockade.
    }
    \label{fig:fig1}
\end{figure}

\section{Photon blockade}

\input{blockade}

\begin{figure*}
    \centering
    \includegraphics[width=0.9\textwidth]{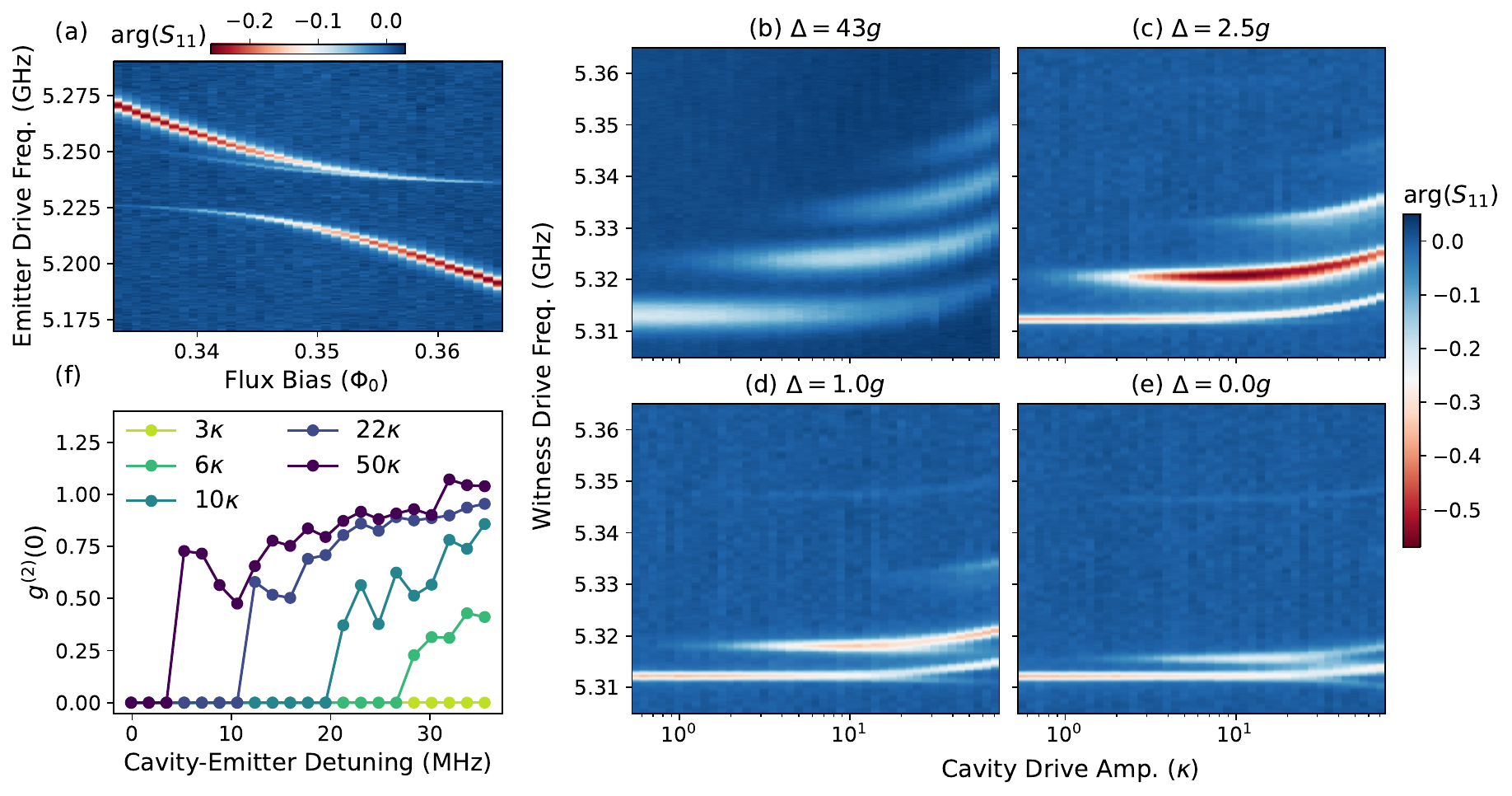}
    \caption{Photon blockade in a Jaynes-Cummings system with strong coupling. 
    (a) By varying the external magnetic flux the emitter transmon is tuned through an avoided crossing with the cavity from which we can extract the coupling $g=13.7~\mathrm{MHz}$ and the cavity-emitter polariton frequencies at various detunings, $\Delta$. Photon number resolving spectroscopy is performed at cavity-emitter detunings 
    (b) $\Delta=43g$, 
    (c) $\Delta=2.5g$, 
    (d) $\Delta=g$, and 
    (e) $\Delta=0$ and drive amplitudes ranging from $\eta < \kappa$ to $\eta \gg \kappa$. Acquiring data for each of the four datasets took 24 minutes.
    (f) As discussed in the text, these PNR spectra are used to estimate $g^{(2)}(0)$ as a function of cavity-emitter detuning. The different colored points represent different cavity drive amplitudes which are specified in the legend. 
    }
    \label{fig:fig3}
\end{figure*}

\section{Photon number resolving spectroscopy}

\input{pns}

\section{Experimental results}

\input{experiment}

\begin{figure}[htp]
    \centering
    \includegraphics[width=\columnwidth]{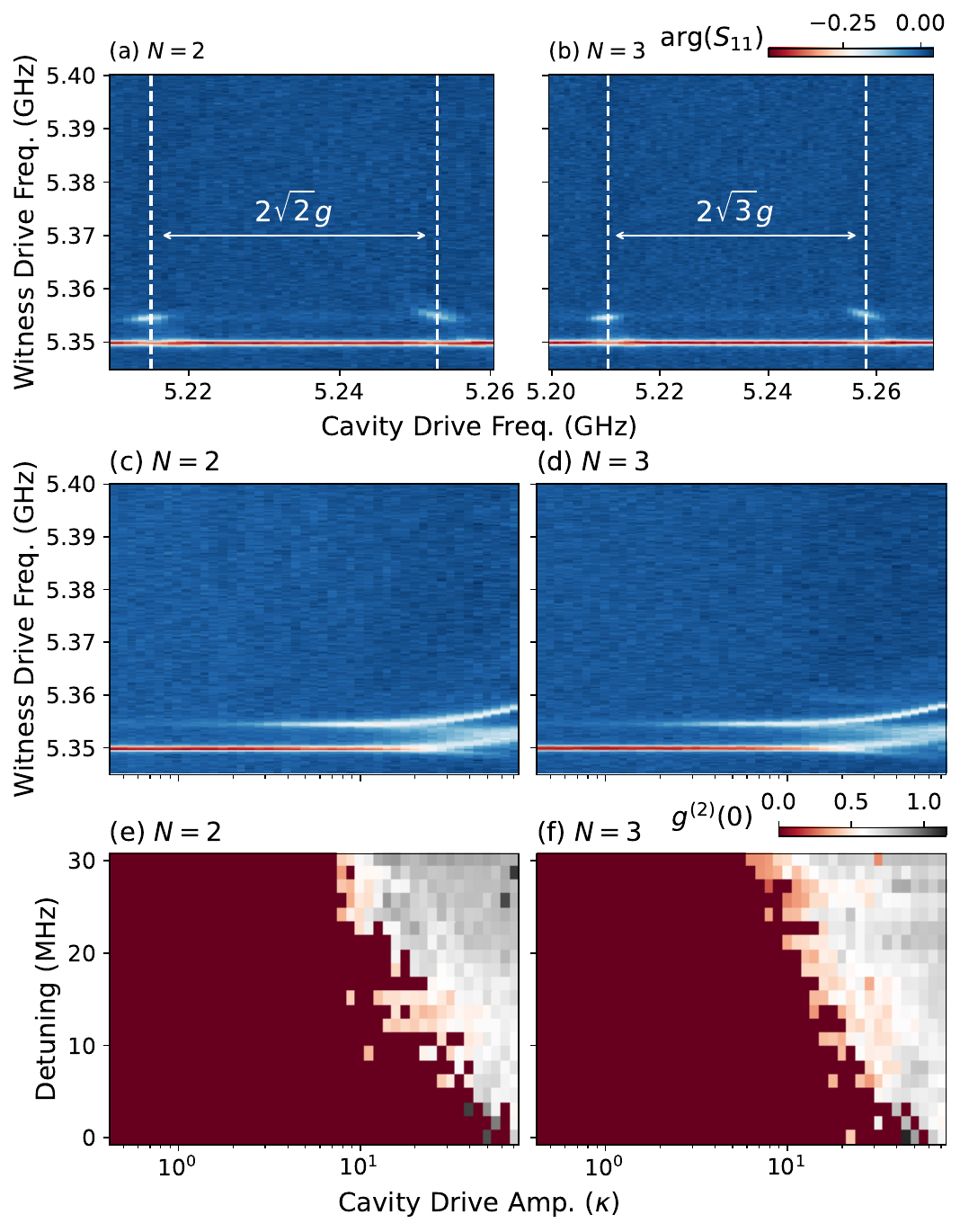}
    \caption{Observation of photon blockade in Tavis-Cummings systems. PNR spectrum with (a) $N=2$ and (b) $N=3$ emitter transmons tuned into resonance with the cavity. The spectrum is measured with fixed cavity drive amplitude $\eta=10\kappa$ while sweeping the cavity drive frequency across the polariton frequencies. Vertical dashed lines mark the polariton frequencies with splitting $2g_{\mathrm{col}} \approx 2\sqrt{N}\bar{g}$. The PNR spectrum for (c) $N=2$ and (d) $N=3$ as a function of the drive amplitude with drive frequency fixed at the lower polariton frequency. The $g^{(2)}(0)$ is extracted for (e) $N=2$ and (f) $N=3$ by repeating the above measurements at various cavity-emitter detunings.
    }
    \label{fig:fig2}
\end{figure}

\section{Discussion}

\input{discussion}

\section{Acknowledgments}
Authors acknowledge support by Noyce Foundation, National Science Foundation CAREER program (Award 2047564), the UC Multicampus Research Programs and Initiatives of the University of California (Grant Number M23PL5936).
This material was funded in part by the U.S. Department of Energy, Office of Science, Office of Advanced Scientific Computing Research Quantum Testbed Program under contracts DE-AC02-05CH11231.
This paper describes objective technical results and analysis.
Any subjective views or opinions that might be expressed in the paper do not necessarily represent the views of the U.S. Department of Energy or the United States Government.

\nocite{*}

\bibliography{references}

\newpage 
\clearpage
\pagebreak

\onecolumngrid
\begin{center}
\textbf{\large Supplementary Information for: ``Observation of Photon Blockade in a Tavis-Cummings System''}
\end{center}

\input{supplement-in-main}

\end{document}

%% file: intro.tex
Photon-photon interactions are notoriously absent in nature except within media that exhibit optical nonlinearity -- the phenomenology explored in classical and quantum optics alike. At the end of the 20$^\text{th}$ century, scientists studied Kerr and Kerr-like nonlinearities for applications in photon logic and single-photon turnstile devices. Among these works was the theory of \emph{photon blockade} (PB), where the absorption of one photon blocks the transmission of a subsequent photon \cite{PhysRevLett.79.1467}. Its experimental realizations in atomic, quantum dot and superconducting Kerr and Jaynes-Cummings systems followed \cite{Birnbaum2005, Faraon2008, PhysRevLett.107.053602, Lang2011}. Next came observations of the unconventional photon blockade in solid state systems, which eased the conditions for demonstrating PB effect toward weakly (as opposed to strongly) interacting oscillators \cite{PhysRevA.96.053810, Vaneph2018}. Photon blockade has recently been proposed for Tavis-Cummings systems \cite{PhysRevA.96.011801, trivedi2019photon}, which are of particular relevance for defect-related quantum emitters in semiconductors where strong cavity quantum electrodynamical (QED) coupling is currently achieved only with emitter ensembles \cite{Janitz:20, 10.1063/5.0130196, 10.1063/5.0077045}.

Photon blockade is typically manifested in terms of the statistics of photons emitted from the cavity, where the two-photon correlation function $g^{(2)}(\tau) < 1$ is used as an indicator of blockade. 
The correlation function can be measured with a Hanbury Brown-Twiss interferometer using a pair of photon detectors, which is especially straightforward in the optical domain where detectors are readily available and highly efficient \cite{STEVENS201325}.
In the microwave domain where photon detectors are less developed, correlation functions can be determined from quadrature measurements of the field leaking out of the cavity using homodyne or heterodyne detection \cite{daSilva2010, Lang2011, Vaneph2018}. 
These measurements require high quality quantum-limited amplifiers to achieve high SNR and can be very slow if high order moments of the field distribution need to be measured. 
Recent work has used the more direct approach of coupling an auxiliary qubit to the cavity to perform Wigner tomography of the cavity state to verify photon blockade \cite{bretheau2015blockade,Chakram2022}. 
If the qubit and cavity are coupled in the strong dispersive limit the cavity photon number information can be mapped onto the qubit frequency.

In this work we use $N=1,2,3$ superconducting transmon qubits capacitively coupled to a co-planar waveguide (CPW) cavity \cite{blais2004cqed} to achieve the photon blockade condition, observed in the cavity photon number. 
An additional transmon qubit in the strong-dispersive regime is used as a probe of the distribution of photon number states in the cavity. 
We employ photon number resolving (PNR) spectroscopy \cite{gambetta_2006, schuster_2007} of this ancillary qubit to detect which photon number states of the cavity are occupied in the presence and absence of photon blockade. 
We show that as the blockading qubits are tuned into resonance with the cavity, the cavity occupation transitions from approximately Poissonian with up to $n=4$ photons to sub-Poissonian occupation limited to only $n=0,1$. Our theoretical predictions explain the experimental data closely.
These measurements constitute a simpler and more natural method for detection of photon blockade in superconducting cavity QED systems in the microwave regime.

%% file: blockade.tex
A cavity mode with frequency $\omega_{\mathrm{c}}$ coupled to $N$ two-level quantum emitters with frequency $\omega_{\mathrm{a}}$ is described by the Tavis-Cummings (TC) Hamiltonian in the Rotating Wave Approximation ($\hbar=1$ here and throughout):
\begin{equation}
H_\text{TC} = \omega_{\mathrm{c}} a^\dagger a + \sum_i^N \left( \omega_{\mathrm{a}} \sigma_i^+\sigma_i^- + g(a^\dagger \sigma_i^- + a \sigma_i^+) \right),
\label{eq:HTC}
\end{equation}
where $a$ ($a^\dagger$) is the lowering (raising) operator for the cavity mode, $\sigma_i^-$ ($\sigma_i^+$) is the lowering (raising) operator for the $i^\text{th}$ emitter, and $g$ is the cavity-emitter coupling which we assume to be identical for all $N$ emitters.

In the strong-coupling regime, the eigenstates of $H$ are mixtures of light and matter known as polaritons.
When the system is driven on resonance with a polariton in the $n=1$ subspace, the drive field is detuned from all 1-to-2 excitation transitions.
Considering a coherent drive at the frequency of the lowest polariton ($\omega_d=\omega_{\mathrm{c}}-\sqrt{N}g$), the detuning between $\omega_d$ and the transition to the nearest $n=2$ state is $(2\sqrt{N} - \sqrt{4N-2})g$.
If this detuning is large relative to $\kappa$ (the cavity linewidth) and $\gamma$ (the emitter linewidth), then the drive fails to significantly couple the $n=1$ polariton to any state of the $n=2$ manifold (see Fig.~\ref{fig:fig1}d).

%% file: pns.tex
The photon number statistics of the blockaded cavity can be extracted from the spectral response of a dispersively coupled qubit. This can be understood by again considering the Hamiltonian in Eqn.~\ref{eq:HTC} with $N=1$ emitter, now termed the witness qubit, and working in the dispersive limit $g \ll \Delta_{\mathrm{w}}=\omega_{\mathrm{w}}-\omega_{\mathrm{c}}$ where $\omega_{\mathrm{w}}$ is the witness qubit frequency. A Schrieffer-Wolff transformation of that Hamiltonian yields
\begin{equation}
    H_{\mathrm{disp}}=\omega_{\mathrm{c}} a^{\dagger}a+\left(\tilde{\omega}_{\mathrm{w}}+2\chi a^{\dagger}a\right)\sigma^{+}\sigma^{-}
\end{equation}
where $\tilde{\omega}_{\mathrm{w}}=\omega_{\mathrm{w}}+g^{2}/\Delta_{\mathrm{w}}$ is the Lamb-shifted emitter frequency and
\begin{equation}
    \chi=\frac{g^{2}}{\Delta_{\mathrm{w}}\left(1-\frac{\Delta_{\mathrm{w}}}{\alpha}\right)}
\end{equation}
for the transmon which is modified by the presence of higher levels from the usual $\chi_{\mathrm{TLS}}=g^{2}/\Delta_{\mathrm{w}}$ for a pure two-level system (TLS). The interaction term can be viewed as a photon number dependent shift of the transmon frequency by $2\chi$ for every additional photon. The strong-dispersive regime is characterized by $\chi>\kappa,\gamma$ with $\kappa$ and $\gamma$ the cavity and qubit decay rates respectively. In this limit it has been predicted \cite{gambetta_2006} and observed \cite{schuster_2007} that the transmon response spectrum $S(\omega)$ will split into multiple resolvable peaks, separated by $2\chi$, corresponding to different numbers of photons in the cavity. The weights of the various peaks are approximately proportional to the probability of the corresponding photon number in the cavity. Therefore, measuring the qubit spectrum gives an estimate of the photon number distribution in the cavity state.

%% file: experiment.tex
We implement the TC model in a system consisting of eight flux-tunable transmon qubits capacitively coupled to a CPW resonator. The device is introduced and described in detail in Ref.~\cite{marinelli_2023}. The cavity frequency is $\omega=5.230~\mathrm{GHz}$, with cavity-emitter couplings $g=12.5-13.7~\mathrm{MHz}$ ($\bar{g}=13.2$ MHz), emitter decay rates $\gamma \sim 0.1~\mathrm{MHz}$, and cavity linewidth $\kappa=0.1~\mathrm{MHz}$. The coupling rates are extracted from fitting avoided crossings between the transmons and the cavity, like the one shown in Fig.~\ref{fig:fig3}a. These avoided crossings are measured by using an on-chip flux line to tune the transmon frequency through the cavity frequency while performing two-tone spectroscopy on the transmon. This demonstrates that the transmons and cavity are in the strong coupling regime ($g \gg \kappa,\gamma$).

We begin by showing that PNR spectroscopy is an effective probe of photon blockade by applying it to the Jaynes-Cummings system with $N=1$ emitter blockading the cavity. Photon blockade in JC systems has been studied extensively in the past and serves as a benchmark of the method. 

The witness transmon, $Q_{\mathrm{w}}$, is tuned to $\omega_{\mathrm{w}}=5.313~\text{GHz}$ and has coupling to the cavity $g_{\mathrm{w}}=17~\text{MHz}$, detuning from the cavity $\Delta_{\mathrm{w}}=\omega_{\mathrm{w}}-\omega_{\mathrm{c}}=83~\text{MHz}$, and anharmonicity $\alpha=227~\text{MHz}$. This leads to a dispersive shift of $\chi=5.5~\text{MHz}$ which is easily resolved spectroscopically. A vector network analyzer (VNA) is used to apply a weak readout tone to the readout resonator of the witness qubit and measure the relative amplitude and phase of the reflected signal ($S_{11}$), as a monitor of the qubit state. A weak probe tone is applied to the witness qubit charge drive line and swept across the bare qubit frequency and dispersively shifted qubit frequencies corresponding to up to $n=4$ photons at $\omega_{\mathrm{w}}+8\chi$ to probe the qubit spectral response.

In the empty cavity limit where the emitter transmon is far off resonant with the cavity, $\Delta=43g$, a drive resonant with the cavity frequency, $\omega_{\mathrm{d}}=\omega_{\mathrm{c}}$ is applied to the cavity drive line with varying amplitude. The resulting qubit spectrum is shown in Fig.~\ref{fig:fig3}b. At very low powers a single peak in the spectrum is observed, corresponding to the bare qubit frequency. As the drive amplitude is increased the cavity is populated with non-zero average photon number, $\braket{a^{\dagger}a} > 0$, in steady-state and the qubit spectrum shows additional peaks with relative shifts $2\chi$ corresponding to the dispersively shifted qubit frequency for different numbers of photons in the cavity. The relative weights of the peaks give an approximation of the photon number distribution in the steady state at each drive amplitude. 

Since $g_{\mathrm{w}}/\Delta_{\mathrm{w}} \approx 0.20$, there is some residual hybridization between the cavity and witness qubit which results in the measured photon number distribution deviating from the expected Poisson distribution of a linear cavity excited by a classical drive. Despite this, peaks corresponding to photon numbers up to $n=4$ are clearly observed. In Supp. Sec.~S5 we compare the experimental results to simulations including and excluding the witness qubit to show that the presence of the witness qubit has a non-qualitative effect on the photon blockade.

We repeat the PNR spectroscopy measurement at various detunings, $\Delta$, between an emitter transmon $Q_{1}$ and the cavity to show the emergence of the photon blockade effect as the emitter transmon is brought into resonance with the cavity. The results are shown in Fig.~\ref{fig:fig3}c-e. At each detuning, the cavity is driven at the frequency of the cavity-like polariton (determined from the fit to the avoided crossing). We observe two important changes to the PNR spectrum of the witness qubit as the emitter-cavity detuning is reduced. First, the higher ($n>1$) photon lines gradually fade and disappear until only the $n=0,1$ spectral features are left when the emitter and cavity are resonant ($\Delta=0$) and the cavity photon number is blockaded. Further, we observe a reduction in the spacing between the $n=0$ and $n=1$ photon lines from $2\chi$ to $\chi$. This can be understood by noting that the polariton states have an average of $1/2$ photon when the emitter and cavity are resonant, $\braket{\pm|a^{\dagger}a|\pm}=1/2$ so the resulting dispersive shift on the witness qubit when they are excited is halved. In general, the shift is approximately proportional to the weight of the cavity-like part of the polariton at each detuning. More details can be found in Supp. Sec.~S1. 

We also note a feature in the witness qubit spectrum near 5.35~GHz that moves down slightly as the cavity-emitter detuning is reduced. While we have not determined the exact origin of this feature we have ruled out that it is due to the multi-level nature of the emitter qubits or the presence of the witness qubit.  

Photon blockade is typically demonstrated by a low value ($< 1$) of the two-photon correlation function 
\begin{equation}
g^{(2)}(0) = \frac{\langle a^\dagger a^\dagger a a\rangle}{\langle a^\dagger a \rangle^2}.
\label{eq:g2}
\end{equation}
By estimating the cavity occupation distribution from the PNR spectrum we can compute $g^{(2)}(0)$ (see Supp. Sec.~S5 for more details) at each emitter-cavity detuning, shown in Fig.~\ref{fig:fig3}f, at a variety of cavity drive amplitudes. We observe two important features of photon blockade in JC systems. As the drive strength is increased, the photon blockade gradually weakens as the driving rate overcomes the detuning between the polariton frequency and the frequencies of 1-to-2 excitation transitions in the system. Consequently, as the emitter is tuned off-resonant with the cavity this detuning is reduced and the blockade breaks down at successively lower drive strengths. 

Having validated the PNR technique in the case of a JC system we now apply it to Tavis-Cummings systems with $N=2$ and $N=3$ emitter transmons. We begin by probing the dependence of the collective coupling on the number of emitters in the TC system. All emitters are tuned into resonance with the cavity ($\Delta=0$) and we fix the cavity drive amplitude at $\eta=10\kappa$ while sweeping the cavity drive frequency across the polariton frequencies at $\omega_{\mathrm{c}} \pm \sqrt{N}g$. The resulting PNR spectra in Fig.~\ref{fig:fig2}a-b reveal $n=1$ photon features when the cavity is driven near the polariton frequencies and the measured polariton splitting follows the expected $\sqrt{N}$ dependence \cite{fink2009tavis}. 

Next we fix the cavity drive frequency at the lower polariton frequency and sweep the cavity drive amplitude (similar to the measurement in Fig.~\ref{fig:fig3}e). Only the $n=0$ and $n=1$ photon features are observed in the PNR spectra (Fig.~\ref{fig:fig2} c-d), thus demonstrating the observation of photon blockade in the TC systems with both $N=2$ and $N=3$ emitters. 

Again, we extract $g^{(2)}(0)$ across a wide range of cavity-emitter detunings and cavity drive amplitudes, observing a similar dependence as the JC case (Fig.~\ref{fig:fig2}e-f). As expected, the photon blockade breaks down at large detunings and large drive amplitudes where the drive is able to excite the system beyond the $n=1$ excitation manifold \cite{fink2017breakdown}.

%% file: discussion.tex
Non-classical states of light, particularly single photons, are essential resources for technologies such as quantum communication, quantum computation \cite{Couteau2023}, and quantum sensing \cite{Pirandola2018}. The ability to generate and manipulate these states with high fidelity and efficiency is a major hurdle to realizing the full potential of quantum technologies. Photon blockade presents one well-known avenue for imprinting the inherent non-classicality of a two-level quantum system onto an optical mode.

In this work, we have successfully used photon number-resolving (PNR) spectroscopy to observe photon blockade in a circuit QED system containing up to $N=3$ transmon qubits. This is confirmed by the vanishing of the PNR peaks for photon numbers greater than one when $N=1,2,3$ transmons are tuned onto resonance with the cavity for blockade, and by the $\sqrt{N}$ scaling of the separation between the polaritons in the single-excitation manifold.

While photon blockade has traditionally been studied in the optical domain as a route towards generating non-classical light and potentially realizing on-demand single-photon sources, the high degree of control and flexibility afforded by superconducting circuits makes them an attractive platform for exploring blockade physics. The PNR spectrum of the witness qubit at a particular cavity-emitter detuning and cavity drive amplitude is measured in 32 seconds. Multiple $n^{\mathrm{th}}$ order correlations can be calculated from this data quickly in postprocessing following the  Eq.~(19) in Supp. Sec.~5. In comparison, similar measurements in quantum dot-cavity Jaynes-Cummings systems take seconds, minutes and hours for $g^{(2)}$, $g^{(3)}$ and $g^{(4)}$ correlations, respectively \cite{rundquist2014nonclassical}. Thus, this technique offers shorter measurement times when compared to the optical systems using photo-detectors. These faster measurement times come at the expense of reduced sensitivity to very small values of the correlation function, $g^{(2)}(0) \ll 1$. The ability to engineer strong photon-photon interactions and tune system parameters in situ may allow for the investigation of not only conventional photon blockade but also more exotic regimes such as unconventional \cite{PhysRevA.96.053810} and subradiant blockade \cite{PhysRevA.96.011801}. Furthermore, the strong dispersive coupling easily achieved in these systems, combined with the ability to directly probe the ``witness" qubit, circumvents the need for highly efficient single-photon detectors required in optical experiments, which are not available at microwave energies.

While quantum communication and networking protocols have been extensively explored using optical photons, the development of microwave-based quantum networks could offer several advantages. These include compatibility with existing telecommunications infrastructure and the potential for seamless integration with superconducting quantum processors. Recent progress in microwave-to-optical photon conversion \cite{Lauk2020} and the development of low-loss microwave quantum channels \cite{Axline2018} suggest that superconducting systems could play a significant role in future hybrid quantum networks, leveraging the strengths of both microwave and optical domains.

The successful observation of photon blockade using PNR spectroscopy in superconducting circuits also opens up new possibilities for studying and comparing blockade physics across different platforms. While optical and superconducting systems have distinct advantages and challenges, there may be valuable insights to be gained from comparing the two. For example, the ability to engineer and control interactions in superconducting architectures could provide a means to simulate and study blockade phenomena that are challenging to access in optical systems. By comparing the blockade behavior and its dependence on system parameters across these different platforms, we may gain a deeper understanding of the fundamental physics and identify new avenues for enhancing the performance of blockade-based single-photon sources in both optical and microwave domains.

All datasets, QuTiP simulation code and python analysis notebooks are available at \url{github.com/radulaski/PhotonBlockadeTavisCummings}.

%% file: supplement-in-main.tex
\renewcommand{\thesection}{S\arabic{section}}
\renewcommand{\thefigure}{S\arabic{figure}}
\renewcommand{\thetable}{S\arabic{table}}
\setcounter{table}{0}
\setcounter{section}{0}
\setcounter{figure}{0}   
\renewcommand{\theequation}{S\arabic{equation}}
\setcounter{equation}{0}   

\section{Effective Dispersive Shift}

\input{supp-effective_chi}

\section{Drive Amplitude Calibration}

\input{supp-drive_strength_cal}

\section{Experimental Setup}

\input{supp-wiring}

\section{Numerical simulation with QuTiP}

\input{supp-qutip_simulation}

\section{Extracting $g^{(2)}(0)$ from PNR Spectra}

\input{supp-extract-g2}

%% file: supp-effective_chi.tex
We derive analytically the dispersive shift of the witness qubit for each occupation state of the resonant cavity with blockade qubit(s).

\subsection{$N=1$ blockade qubit}
We consider the case of $N=1$ blockade qubits first as the simplest example.
There are two cases to consider: $n=1$ and $n>1$ excitations.
The $n=1$ excited manifold is spanned by the basis $\{|1,g,g\rangle, |0,e,g\rangle, |0, g, e\rangle\}$, where the three entries of each state label the cavity photon-number, the blockade qubit state, and the witness qubit state, respectively.
In this basis, the Hamiltonian reads
$$
H = \begin{pmatrix}
\omega_c-\chi & g & 0 \\
g & \omega_a & 0 \\
0 & 0 & \omega_w+\chi
\end{pmatrix}.
$$
The upper left block corresponds to the subspace where the witness qubit is in $|g\rangle$.
The eigenstates of this block are
$$
|\psi_{N=1,n=1}^{g\pm}\rangle = -\frac{1}{2g} \left( \Delta + \chi \pm \sqrt{4g^2 + (\Delta + \chi)^2} \right) |1,g,g\rangle + |0,e,g\rangle,
$$
where $\Delta \equiv \omega_a - \omega_c$ and we have ignored normalization for readability.
These states have energies
$$
E_{N=1,n=1}^{g\pm} = \frac{1}{2} \left( \omega_a + \omega_c - \chi \pm \sqrt{4g^2 + (\Delta + \chi)^2} \right),
$$
and correspond to the familiar $n=1$ polaritons of the Jaynes-Cummings Hamiltonian, plus a dispersive shift.

The lower right block of the $n=1$ Hamiltonian corresponds to the single excitation residing in the witness qubit:
$$
	|\psi_{N=1,n=1}^{e}\rangle = |0,g,e\rangle,
$$
which has energy
$$
	E_{N=1,n=1}^e = \omega_w + \chi.
$$

In the $n>1$ excitation case, using the basis $\{ |n,g,g\rangle, |n-1,e,g\rangle, |n-1,g,e\rangle, |n-2,e,e\rangle \}$, the Hamiltonian reads
$$
H = \begin{pmatrix}
	n(\omega_c-\chi) & g\sqrt{n} & 0 & 0 \\
	g\sqrt{n} & (n-1)(\omega_c - \chi) + \omega_a & 0 & 0 \\
	0 & 0 & (n-1)(\omega_c+\chi) + \omega_w + \chi & g\sqrt{n-1} \\
	0 & 0 & g\sqrt{n-1} & (n-2)(\omega_c+\chi) + \omega_w + \chi + \omega_a
\end{pmatrix}.
$$
Again, the upper left block corresponds to states analogous to $n$-excitation Jaynes-Cummings polaritons (again ignoring normalization):
$$
	|\psi_{N=1,n>1}^{g\pm}\rangle = -\frac{1}{2\sqrt{n}g} \left( \Delta + \chi \pm \sqrt{4g^2n + (\Delta + \chi)^2} \right) |n,g,g\rangle + |n-1,e,g\rangle,
$$
with corresponding energies
$$
	E_{N=1,n>1}^{g\pm} = \frac{1}{2} \left( \omega_a + (2n-1)(\omega_c - \chi) \pm \sqrt{4g^2n + (\Delta + \chi)^2} \right).
$$
The lower left block also represents states similar to Jaynes-Cummings type polaritons, now with $n-1$ excitations in the polariton part and one excitation in the witness:
$$
|\psi_{N=1,n>1}^{e\pm}\rangle = -\frac{1}{2\sqrt{n-1}g} \left( \Delta + \chi \pm \sqrt{4g^2(n-1) + (\chi - \Delta)^2} \right) |n-1,g,e\rangle + |n-2,e,e\rangle
$$
(where we have again ignored normalization) with corresponding energies:
$$
E_{N=1,n>1}^{e\pm} = \frac{1}{2} \left( \omega_a + (2n-3)\omega_c + 2\omega_w + (2n-1)\chi \pm \sqrt{4g^2(n-1) + (\chi-\Delta)^2} \right).
$$

We now map out the PNR transitions of the witness qubit. From here on we consider the blockade qubits to be on resonance with the cavity so that $\Delta=0$.
Assume the system starts in its ground state $|0,g,g\rangle$. When the witness qubit is probed, if it is flipped to $|e\rangle$, the system moves to $|\psi^e_{N=1,n=1}\rangle$, with energy $\omega_w+\chi$, reflecting the Lamb shift of the witness.
If the system instead starts in an $n$-excitation polariton-like state $|\psi^{g\pm}_{N=1,n>0}\rangle$ where the witness is in $|g\rangle$, then exciting the witness qubit couples the system to states in the $(n+1)$-excitation manifold:

\begin{align}
	\langle \tilde{\psi}_{N=1,n+1}^{e\pm} | \sigma^x_w | \tilde{\psi}_{N=1,n>0}^{g\pm}\rangle &= \frac{2\sqrt{n}g}{\sqrt{4ng^2 + \chi^2}} \approx 1 \label{approx1} \\
	\langle \tilde{\psi}_{N=1,n+1}^{e\mp} | \sigma^x_w | \tilde{\psi}_{N=1,n>0}^{g\pm}\rangle &= \pm\frac{\chi}{\sqrt{4ng^2 + \chi^2}} \approx 0, \label{approx2}
\end{align}

where $\sigma^x_w$ is the Pauli X operator acting on the witness qubit, the tildes indicate normalization, and the approximation is made in the limit $\chi \ll g$.
So when the cavity and blockade qubit contains $n$ excitations, the witness qubit can be flipped when addressed with energy
\begin{equation}
    E_{N=1,n+1}^{e\pm}-E_{N=1,n}^{g\pm} = \omega_w+2n\chi.
\end{equation}

The fact that the witness qubit can be flipped without changing the state of the rest of the system means that we can also obtain the PNR spectrum of the witness by considering the eigenstates of the blockade portion of the system in isolation:
$$
\omega_w + (1 + 2\langle \psi_{N=1,n} | a^\dagger a | \psi_{N=1,n} \rangle) \chi = \omega_w + 2n\chi,
$$
where $| \psi_{N=1,n} \rangle$ is the eigenstate of an ordinary JC system with $n$ excitations (without the witness).

We also note that the small effect of the witness can also be observed by the differences between the eigenenergies with ($E^{g\pm}_{N=1,n}$) and without ($E^{\pm}_{N=1,n}$) the witness qubit.
Using our experimental parameters (given in the main text), when the system is resonant ($\omega_a = \omega_c$, $\Delta=0$) we have
\begin{align}
    \Delta_{\mathrm{w/ witness}} = E^{g\pm}_{N=1,n+1} - E^{g\pm}_{N=1,n>0} &= \omega_c - \chi - \frac{1}{2} \sqrt{4ng^2 + \chi^2} + \frac{1}{2} \sqrt{4g^2(n+1) + \chi^2} \\
    \Delta_{\mathrm{w/out witness}} = E^{\pm}_{N=1,n+1} - E^{\pm}_{N=1,n>0} &= \omega_c - \frac{1}{2} \sqrt{4ng^2} + \frac{1}{2} \sqrt{4g^2(n+1)} \\
\end{align}
so that
\begin{equation}
    \Delta_{\text{w/out witness}} - \Delta_{\text{w/ witness}} = \chi + g(\sqrt{n+1} - \sqrt{n}) + \frac{1}{2} \sqrt{4g^2n + \chi^2} - \frac{1}{2} \sqrt{4g^2(n+1) + \chi^2} \approx \chi,
\end{equation}
where the final approximation holds when $\chi \ll \sqrt{n}g$, as is true of our system.
Thus, the effect of the witness on the blockade system is to reduce the energy required to add an excitation to the system by roughly $\chi$, which for our experimental parameters is about 5.5 MHz, or about $\chi / \omega_c = (5.5 \textrm{MHz}) / (5.230 \textrm{GHz}) \approx 0.1\%$.

\subsection{$N>1$ blockade qubits, $n=1$ excitations}
For $N=\{2,3\}$, a similar (albeit tedious) analysis shows that the approximations~(\ref{approx1}, \ref{approx2}) holds when $\chi \ll \sqrt{N}g$.
We omit this derivation and focus on the resulting PNR shifts of the witness qubit, which we compute directly from the mean photon number of the relevant eigenstates of the blockade system without the witness qubit, as discussed above.

Since we focus only on the case of resonant, identical blockade qubits, we simplify matters by working in the basis of collective states, where $N$ spin-$1/2$ systems are considered to act as a single spin-$N/2$ system.
Regardless of $N$, the basis $\{ |1,-N/2\rangle, |0,-N/2+1\rangle \}$ spans the single-excitation manifold. In this basis we have

\begin{equation}
H = \begin{pmatrix}
	n\omega_c & \sqrt{N}g \\
	\sqrt{N}g & (n-1)\omega_c+\omega_a
\end{pmatrix},
\end{equation}

with polariton-like eigenstates (ignoring normalization):
$$
|\psi_{N>1,n=1}^\pm\rangle = -\frac{1}{2\sqrt{N}g} \left( \Delta + \chi \pm \sqrt{4Ng^2 + (\Delta + \chi)^2} \right) |1,g,g\rangle + |0,e,g\rangle.
$$
Considering the resonant case $\omega_a=\omega_c$, straightforward calculation gives
$$
(1 + 2\langle \tilde{\psi}^\pm_{N>1,n=1} | a^\dagger a | \tilde{\psi}^\pm_{N>1,n=1} \rangle) \chi = 2 \chi,
$$
where the tildes indicate normalization.

\subsection{$N=2$ blockade qubits}
With the $n=1$ case covered by the foregoing, we can immediately move to $n>1$.
$$
H = \begin{pmatrix}
	n\omega_c & \sqrt{2n}g & 0 \\
	\sqrt{2n}g & (n-1)\omega_c + \omega_a & \sqrt{2(n-1)}g \\
	0 & \sqrt{2(n-1)}g & (n-2)\omega_c + 2\omega_a
\end{pmatrix}.
$$
There are now more than two eigenstates; we consider only the highest and lowest (denoted by $\pm$), since these are the states most likely to be involved in multi-photon events when the system is driven on resonance with either of the $n=1$ eigenstates. As before, setting $\omega_c=\omega_a$ we compute:
$$
(1 + 2\langle \tilde{\psi}^\pm_{N=2,n>1} | a^\dagger a | \tilde{\psi}^\pm_{N=2,n>1} \rangle) \chi = \left( 2n -1 + \frac{1}{2n-1} \right) \chi,
$$
which agrees with the $n=1$ case as well.

\subsection{$N=3$ blockade qubits}
We must consider two special cases: $n=2$ and $n>2$.

\subsubsection{$n=2$ excitations}
$$
H = \begin{pmatrix}
	2\omega_c & \sqrt{6}g & 0 \\
	\sqrt{6}g & \omega_c+\omega_a & 2g \\
	0 & 2g & 2\omega_a
\end{pmatrix}.
$$
Again, setting $\omega_c=\omega_a$, we compute:
$$
(1 + 2\langle \tilde{\psi}_{N=3,n=2} | a^\dagger a | \tilde{\psi}_{N=3,n=2} \rangle) \chi = 3.2\chi
$$

\subsubsection{$n>2$ excitations}
$$
H = \begin{pmatrix}
	n\omega_c & \sqrt{3n}g & 0 & 0 \\
	\sqrt{3n}g & (n-1)\omega_c+\omega_a & 2\sqrt{n-1}g & 0 \\
	0 & 2\sqrt{n-1}g & (n-2\omega_c)+2\omega_c & \sqrt{3(n-2)}g \\
	0 & 0 & \sqrt{3(n-2)}g & (n-3)\omega_c + 3\omega_a	
\end{pmatrix}.
$$
Again, setting $\omega_c=\omega_a$, we compute:
$$
(1 + 2\langle \tilde{\psi}^\pm_{N=3,n>2} | a^\dagger a | \tilde{\psi}^\pm_{N=3,n>2} \rangle) \chi = \left( 2n - 2 + \frac{6}{\sqrt{16n(n-2) + 25}} \right)\chi,
$$
which agrees with the $n=\{1,2\}$ cases as well.

\subsection{$N > 1$ blockade qubits, $n=2$ excitations}
When blockade is in effect, only the $n=0$ and $n=1$ lines will appear in the PNR spectrum.
The positions of both those lines as a function of $N$ in the general case are given in the foregoing.
We now give the position of the $n=2$ line (the first line suppressed by blockade) as a function of $N$.
\begin{equation}
    H = \begin{pmatrix}
        2\omega_c & \sqrt{2n}g & 0 \\
        \sqrt{2n}g & \omega_c+\omega_a & \sqrt{2n-2}g \\
        0 & \sqrt{2n-2}g & 2\omega_a
    \end{pmatrix}
\end{equation}
On resonance ($\omega_a=\omega_c$) we find for the upper and lower states $\psi_{N>1,n=2}^\pm$,
\begin{equation}
    (1 + 2 \langle \psi_{N>1,n=2}^\pm | a^\dagger a | \psi_{N>1,n=2}^\pm\rangle) \chi = \frac{6N-1}{2N-1}\chi
\end{equation}
so that the shift from the $n=1$ level to the $n=2$ level is
\begin{equation}
    \frac{6N-1}{2N-1}\chi - 2\chi = \frac{2N}{2N-1}\chi,
\end{equation}
which agrees with the $N=1$ case as well.

\subsection{Intuition}
These results can be understood as follows.
The first excitation of the cavity + blockade qubit(s) is exactly half cavity-like regardless of $N$, because there is only one way to split a single excitation between the cavity and the qubit(s).
Therefore populating this state shifts the witness qubit by $\chi$.
When a second excitation is added to the cavity + blockade qubit(s), the number of ways to distribute this excitation depends on the number of blockade qubit(s) $N$.
For $N=1$, the blockade qubit is already saturated and so each additional excitation is purely cavity-like, shifting the witness a further $2\chi$.
For small $N>1$, the blockade qubits are partially but not completely filled by the first excitation, and so the second excitation is partly in the cavity, partle in the qubits.
For large TC systems ($N\gg 1$) the collective state space of the blockade qubits becomes effectively harmonic, and so each additional excitation is equally shared between cavity and blockade qubit(s), so the dispersive shift of the witness tends to $\chi$ per excitation as $N\to\infty$.
For any finite $N$, as $n\to\infty$ the blockade qubits become saturated, and each additional excitation asymptotically resides purely in the cavity, so subsequent dispersive shifts of the witness tend toward $2\chi$.

%% file: supp-drive_strength_cal.tex
In this section we describe the protocol used to calibrate the ratio between the source drive amplitude in volts (V) and the drive rate $\eta$ of the cavity. As described in the main text, when a qubit is tuned on resonance with the cavity there is a conventional photon blockade effect, preventing the system from being excited to beyond the single-excitation manifold spanned by the polariton states $\ket{\pm}$. For sufficiently small drive amplitudes, $\eta \ll g$, we can isolate the vacuum state and one of the polaritons to form a two-level system (TLS), $\{\ket{0},\ket{\pm}\}$. Then we note that 
\begin{equation}
    \braket{0|\eta(a+a^{\dagger})|-}=\braket{0|\eta(a+a^{\dagger})|+}=\frac{1}{2}\eta
\end{equation}
or more generally if the qubit and cavity are nearly on resonance, $0 < \Delta \ll g$ we still have
\begin{equation}
    \braket{0|\eta(a+a^{\dagger})|-}+\braket{0|\eta(a+a^{\dagger})|+}=\eta.
\end{equation}
Thus, for a given drive voltage amplitude, $V$, we can extract $\eta(V)$ from the Rabi oscillation rates of the TLS spanned by the vacuum and the polaritons under resonant driving. We tune $Q_{5}$ onto resonance with the cavity and drive the system resonantly with each polariton frequency while sweeping the drive amplitude. The resulting Rabi rates are extracted from fitting the qubit population oscillations to a sinusoid with a decaying envelope to account for decoherence. 

\begin{figure}[htp]
    \centering
    \includegraphics[width=0.8\columnwidth]{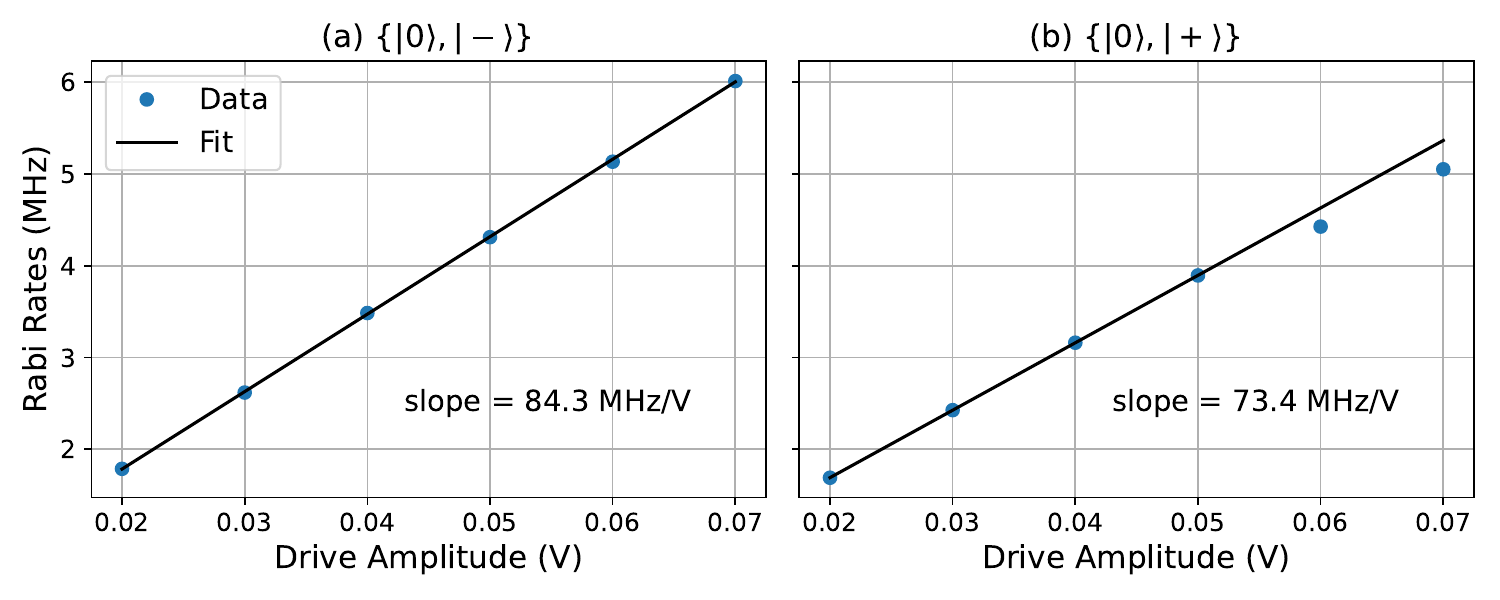}
    \caption{Lower (a) and upper (b) polariton Rabi rates vs. drive amplitude. Rabi rates (blue dots) are extracted from fitting experimentally measured $Q_{5}$ population oscillations while resonantly driving at the polariton frequencies. Linear fit (black line) gives the Rabi rate to drive amplitude ratio in MHz/V.}
    \label{fig:figs1}
\end{figure}

The extracted Rabi rates for both polaritons are shown in Fig.~\ref{fig:figs1}. The linear fit used to extract the drive strength to voltage amplitude ratio uses the first four data points and excludes the highest amplitude results which deviate from the linear trend.

%% file: supp-wiring.tex
\begin{figure}[htp]
    \centering
    \includegraphics[width=0.8\columnwidth]{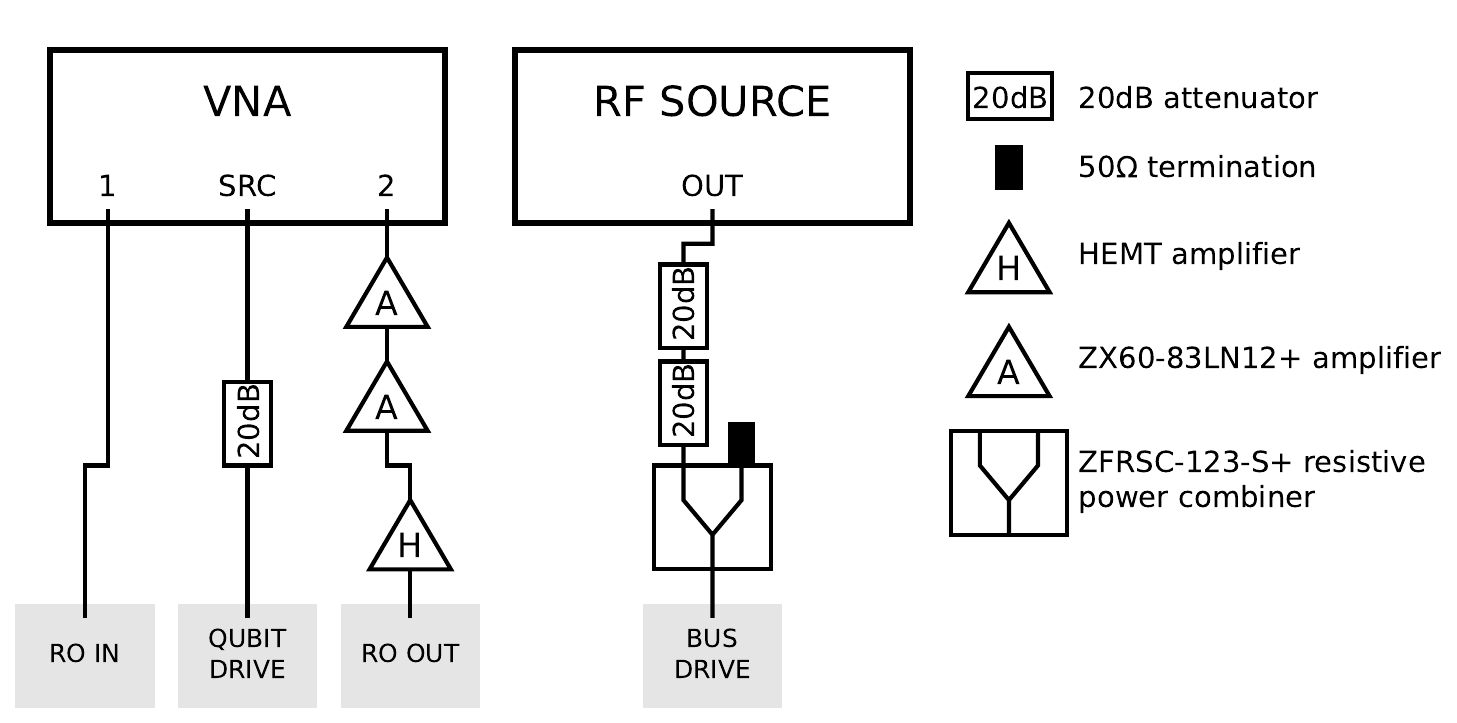}
    \caption{Room temperature wiring diagram for measurements throughout the main text.}
    \label{fig:figs2}
\end{figure}

In Fig.~\ref{fig:figs2} we show the room temperature microwave setup used for the measurements throughout the main text. Details of the cryogenic wiring can be found in the Supplementary Information of Ref.~\cite{marinelli_2023}. A Keysight N5241B Vector Network Analyzer (``VNA'') is used to continuously measure the readout signal, $S_{11}$, with a weak probe tone applied from Port 1 and the reflected signal returned at Port 2. The extra RF source, ``SRC'' from the VNA is used to apply a weak drive on the witness qubit to probe its spectral response. A separate ``RF SOURCE'', AnaPico Aspin20G, is used to drive the cavity. The cavity drive passes through a power combiner to enable measurements where the cavity is driven simultaneously by multiple tones (not presented in this work).  

%% file: supp-qutip_simulation.tex
To simulate this situation, we add a coherent drive term
\begin{equation}
    H_\text{drive} = \eta a^\dagger e^{-i \omega_dt} + \text{h.c.}
\end{equation}
to form the Hamiltonian $H = H_\text{TC} + H_\text{drive}$, where $\omega_d$ is the drive frequency and $\eta$ is the drive field amplitude (in units of frequency; photons per unit time).
The time dependence is removed by unitary transformation into a frame co-rotating with the drive tone:
\begin{align*}
\tilde{H} &= U^\dagger H U - iU \frac{dU}{dt} \\
&= \Delta \hat{n} + g\sum_i^N \left( a^\dagger \sigma_i^- + a \sigma_i^+ \right) + \eta(a^\dagger + a)
\end{align*}
where $U = \exp(i\omega_d \hat{n} t)$ and $\Delta\equiv\omega - \omega_d$ is the detuning of the drive.

Loss of excitations to the environment is assumed to be Markovian, so we write a master equation for the overall dynamics of the system in Lindblad form as
\begin{equation}
    \dot{\rho} = -i[\tilde{H}, \rho] + \mathcal{D}_a(\rho) + \sum_i^N \mathcal{D}_{\sigma_i^-} (\rho),
    \label{eq:lindblad}
\end{equation}
where
\begin{equation}
    \mathcal{D}_A (\rho) \equiv A \rho A^\dagger - \frac{1}{2} \left\{A^\dagger A, \rho\right\}_+
\end{equation}
is the incoherent dissipator generated by the operator $A$, so that $\mathcal{D}_a$ and $\mathcal{D}_{\sigma_i^-}$ represent losses to the environment via the cavity and via the $i^\text{th}$ emitter, respectively.

Our experimental measurements represent averages over periods much longer than the characteristic timescales of the system ($\omega, g, \kappa, \gamma$), so we deal exclusively with expectation values and correlation functions of the steady-state ($\dot{\rho}_\text{ss} = 0$), which we obtain by solving Eqn.~\ref{eq:lindblad} using the python package QuTiP.

Photon blockade is typically characterized by low values ($< 0.5$) of the two-photon correlation function
\begin{equation}
    g^{(2)}(0) = \frac{\langle a^\dagger a^\dagger a a \rangle}{\langle a^\dagger a \rangle^2},
\end{equation}
which has the virtue of being straightforwardly implemented in the optical domain in terms of a photon counting experiment.
While this is a measurement on photons emitted from the cavity, in simulation we can compute $g^{(2)}(0)$ using the cavity annihilation operator $a$, since the number of photons emitted from the cavity is proportional to the cavity occupation.

In this work, we use photon-number resolving spectroscopy to observe photon blockade -- directly probing the photon number distribution of the cavity.
In this case, blockade is evidenced by a divergence of the photon-number distribution from the well-known Poisson distribution that represents the steady-state of a bare cavity under coherent drive.
In Fig.~\ref{fig:fig1}(c) we compare an example simulation of the steady-state photon number distributions within two coherently pumped cavities (one bare, one containing a resonant emitter creating a blockade).
The blockaded system shows dramatically attenuated occupation of higher ($>1$) Fock states, and clearly diverges from the Poisson distribution which has the same average photon number.

%% file: supp-extract-g2.tex
\begin{figure}[htp]
    \centering
    \includegraphics[width=\textwidth]{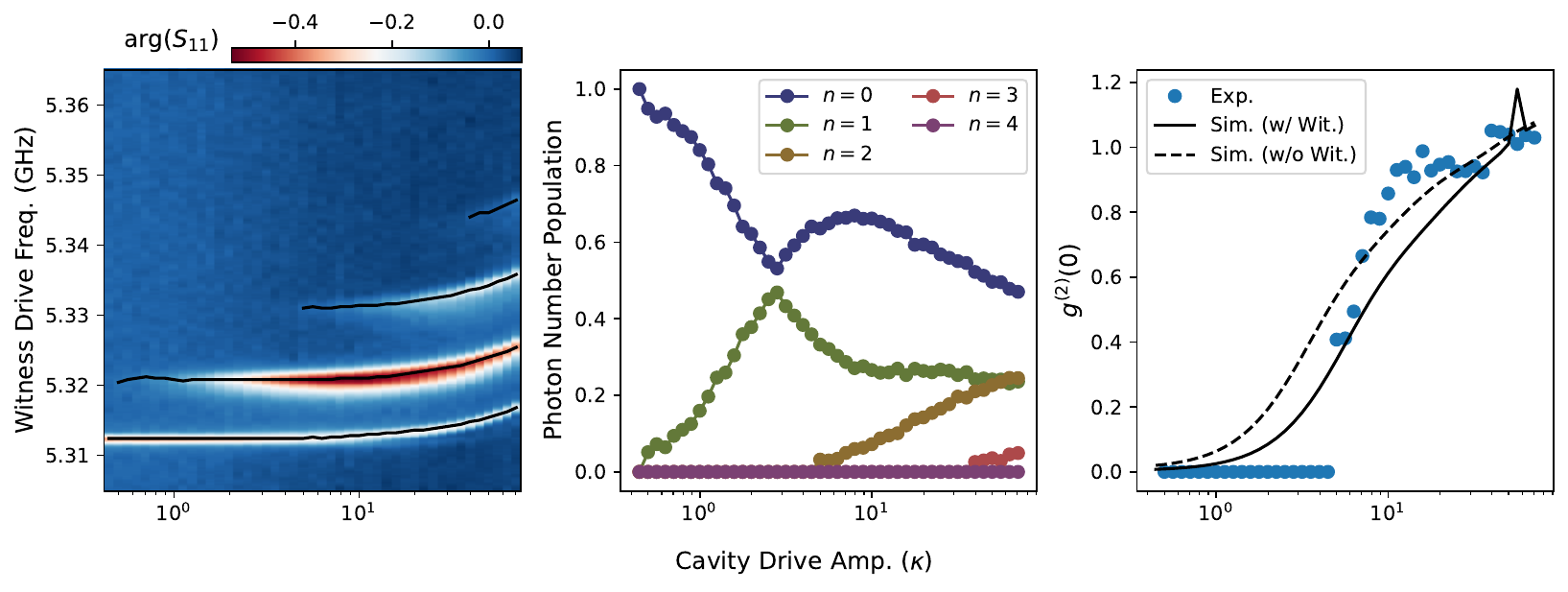}
    \caption{Extracting $g^{(2)}(0)$ from experimental PNR spectra.
    (a) Sample PNR spectrum measured in the same way as Fig.~2b-e in the main text at cavity-emitter detuning $\Delta=2.6g$. Solid black lines overlay the data showing the extracted peak positions of the $n=0,1,2,3$ photon lines.
    (b) The peak prominences for each photon number, $n$, are used as weights and normalized to determine the photon number distribution at each drive amplitude.
    (c) Using the formula in the text the $g^{(2)}(0)$ can be estimated from the measured photon number distribution (blue points). This is compared to master equation simulations of the system's steady state including (solid black line) and excluding (dashed black line) the coupling of the witness qubit to the cavity.
    }
    \label{fig:figsX}
\end{figure}
The second order correlation function used as an indicator of photon blockade,
\begin{equation}
    g^{(2)}(0) = \frac{\langle a^\dagger a^\dagger a a\rangle}{\langle a^\dagger a \rangle^2}
    = \frac{\sum_{n=0}^\infty n(n-1)P(n)}{\left[\sum_{n=0}^\infty nP(n)\right]^2}
    \label{eqn:g2-from-dist}
\end{equation}
can be computed from the photon number distribution in the cavity, $P(n)$. Here we outline the procedure for estimating $P(n)$ from the experimentally measured PNR spectra. First, we find the peaks in the witness qubit spectrum (Fig.~\ref{fig:figsX}a). The peak prominences are taken as the weights, $W_{n}$ of the corresponding photon number $n$ in the distribution (Fig.~\ref{fig:figsX}b). We normalize the distribution so that $P(n)=W_{n}/(\sum_n W_{n})$ and $\sum_{n} P(n)=1$. Finally, we plug these $P(n)$ into Eqn.~\ref{eqn:g2-from-dist} to estimate $g^{(2)}(0)$ at each cavity drive amplitude (Fig.~\ref{fig:figsX}c). This procedure is repeated as we vary the cavity-emitter detunings and the number of emitters, $N$.

Since the PNR technique obtains information about the photon distribution directly, we can in principle also compute higher level correlations:
\begin{equation}
    g^{(m)}(0) = 
    \frac{\sum_{n=0}^\infty \left(\prod_{i=0}^{m-1} (n-i)\right) P(n)}{\left[\sum_{n=0}^\infty nP(n)\right]^m}.
\end{equation}